\documentclass[a4paper,fleqn,usenatbib]{mnras}

\usepackage{newtxtext,newtxmath}
\usepackage{enumerate}

\usepackage[T1]{fontenc}
\usepackage{ae,aecompl}

\usepackage{graphicx}	
\usepackage{amsmath}	
\usepackage{amssymb}	

\newcommand{\Muv}{\ensuremath{M_{\mathrm{UV}}}}
\DeclareRobustCommand{\De}[3]{#2}

\title[Globular Clusters Over Cosmic Time]{Globular Clusters in High-Redshift Dwarf Galaxies: A Case Study from the Local Group}

\author[T. O. Zick]{
Tom O. Zick,$^{1,2}$\thanks{E-mail: tzick@berkeley.edu}
Daniel R. Weisz,$^{1}$
Michael Boylan-Kolchin$^{3}$
\\
$^{1}$Department of Astronomy, University of California Berkeley, Berkeley, CA 94720, USA\\
$^{2}$Lawrence Livermore National Laboratory, PO Box 808 L-210, Livermore, CA, 94551, USA\\
$^{3}$Department of Astronomy, The University of Texas at Austin, 2515 Speedway, Stop C1400, Austin, TX 78712, USA
}

\date{Accepted XXX. Received YYY; in original form ZZZ}

\pubyear{2018}

\begin{document}
\label{firstpage}
\pagerange{\pageref{firstpage}--\pageref{lastpage}}
\maketitle

\begin{abstract}
We present the reconstructed evolution of rest-frame ultra-violet (UV) luminosities of the most massive Milky Way dwarf spheroidal satellite galaxy, Fornax, and its five globular clusters (GCs) across redshift, based on analysis of the stellar fossil record and stellar population synthesis modeling. We find that (1) Fornax's (proto-)GCs can generate $10-100$ times more UV flux than the field population, despite comprising $<\sim 5\%$ of the stellar mass at the relevant redshifts; (2) due to their respective surface brightnesses, it is more likely that faint, compact sources in the Hubble Frontier Fields (HFFs) are GCs hosted by faint galaxies, than faint galaxies themselves. This may significantly complicate the construction of a galaxy UV luminosity function at $z>3$. (3) GC formation can introduce order-of-magnitude errors in abundance matching. We also find that some compact HFF objects are consistent with the reconstructed properties of Fornax's GCs at the same redshifts (e.g., surface brightness, star formation rate), suggesting we may already have detected proto-GCs in the early Universe.  Finally, we discuss the prospects for improving the connections between local GCs and proto-GCs detected in the early Universe.

\end{abstract}




\section{Introduction}
Given their ancient stellar populations and ubiquity, globular clusters (GCs) have long been used as signposts of star formation in the early universe \citep[e.g.,][]{West2004ReconstructingClusters, Brodie2006ExtragalacticFormation, Peng2006TheGalaxies}. This is especially true of the metal-poor GC population for which the average age in the Milky Way is $>12$ Gyr \citep{Vandenberg2013THEISSUES, Forbes2015TheClusters}, corresponding to formation at a redshift of $z>3$, primarily in low mass dwarf galaxies  \citep[e.g.,][]{Searle1978CompositionsHalo, Zinnecker1988TheClusters,Elmegreen2012FORMATIONUNIVERSE, Leaman2013TheGalaxy}. GCs are also uniquely tied to the dark matter (DM) halo masses of their host galaxy \citep{Hudson2014DARKPOPULATIONS, Harris2017GalacticEnvironments} which has been used to infer their number densities at high-redshift \citep{Ricotti2002, Renzini2017, mbk2017theconnection}. Leveraging this to calculate their relative contribution to the high-redshift ultra violet luminosity function (UVLF) has shown GCs may have a non-negligible contribution at absolute UV magnitudes as bright as $\Muv = -17$ and could be easily detectable with the \textit{James Webb Space Telescope (JWST)} under most cluster formation assumptions \citep{Katz2013TwoSatellites, mbk2017theformation}. These projected number densities at high-redshift make proto-GCs compelling in the context of reionization \citep{Ricotti2002, Schaerer2011AReionisation, Katz2013TwoSatellites, mbk2017theformation}.
 
To add to this picture, recent observations may be catching GCs in the act of formation. The Hubble Frontier Fields (HFF) program \citep{Coe2015FrontierResults, lotz2017} which leverages flux amplification due to gravitational lensing of source galaxies by massive foreground galaxy clusters, has allowed investigation of a new faint and compact region of parameter space. Emerging observational constraints on the sizes of faint ($-20 \gtrsim \Muv \gtrsim -12$) galaxies in the HFF at $z \sim 2-8$,  indicate half-light radii $<165$~pc ranging as low as $14$~pc. Sources fainter than $\Muv = -16$ are found to be systematically more compact than originally assumed for completeness estimates in the UVLF \citep{Kawamata2015THEDATA, Laporte2016, Bouwens2017ExtremelyEmissivity, Bouwens2017complexes, kawamata2017}. Recent spectroscopic follow up at $z\sim6$ and $z\sim7$, has tentatively classified some of these compact sources as proto-GCs 
(\citealp{Vanzella2017PavingZ3}, see also \citealt{elmegreen2017} for claims of possible proto-GCs at high-redshift)

While studies in the HFF have pushed measurements of the $z \sim 5-7$ UVLF to as faint as $\Muv = -13$, it remains difficult to disentangle size, completeness and intrinsic magnitude below $\Muv = -15$ \citep{bouwens_uncertainties}. Understanding the sources contributing at faint magnitudes is however necessary to constrain models of reionization, the vast majority of which rely on low mass galaxies \citep[e.g.,][]{Kuhlen2012ConcordanceEvolution, Robertson2013NewCampaign, Robertson2015CosmicTelescope}.

A complementary way to study the faint end of the luminosity function is stellar archaeology \citep{Weisz2014HUBBLE, mbk2015tm, mbk2016}, using resolved stellar populations in the Local Group to reconstruct their star formation histories at high redshift. This has been used to constrain the slope and turnoff of the UVLF at high redshift \citep{mbk2014, Weisz2017LocalEra}. As there are multiple GC-hosting dwarf galaxies in the Local Group (LG) with progenitors relevant to reionization, using the LG as a time machine to study the relative detectability of dwarf host and GCs at high redshift is a viable avenue towards understanding the faint-end UVLF.  

In this work we extend the fossil record approach to reconstruct the intrinsic and observational features of GCs in their dwarf hosts across redshift. To illustrate the potential of this technique we focus on the Fornax dwarf spheroidal and its five GCs \citep[e.g.,][]{McConnachie2012TheGroup, deBoer2012TheGalaxy, Larsen2012ConstraintsDSph, deBoer2016FourDSph}. Though Fornax has a high specific frequency of GCs, this allows us to do the following: 1) study the relative contribution of objects that could feasibly contribute to the faint end of the UVLF and compare their respective detectability; 2) As metal-poor GCs are postulated to form at $z>3$ in low mass galaxies \citep{Searle1978CompositionsHalo, Zinnecker1988TheClusters, Bekki2008TheSimulations, bekki2009, muratov2010, forbes2011, Leaman2013TheGalaxy, tonini2013} and dynamical simulations motivate that Fornax's GCs likely formed \textit{in-situ} \citep{Arca-Sedda2016GlobularCase}, reconstructing their respective star formation histories allows us to study a common avenue for GC formation holding halo mass constant.  

This paper is structured as follows: 
We divide our analysis into a fixed age and a probabilistic approach. In the former we derive the observational signatures of Fornax's GCs assuming complete knowledge of cluster age and in the latter, we investigate the effects of uncertainty in stellar dating by repeating our analysis using a probability distribution function of GC ages. We proceed to place our findings in the context of current high-redshift observations and their physical interpretation. Finally, we discuss the effect of varying the GC birth to present day mass ratio and prospects for connecting local and high-redshift observations. To convert from lookback time to redshift, we adopt a 2016 Plank cosmology \citep{Ade2016iPlanck/iResults}.

\section{Methodology}
In this paper, we use the star formation history (SFH) and ages of globular cluster of Milky Way (MW) satellite galaxy Fornax, combined with stellar population synthesis modeling, to reconstruct their rest frame UV luminosity as a function of redshift. This technique closely follows that described in \citet{Weisz2014HUBBLE} and \citet{mbk2015tm}, and we refer the reader to those papers for complete details. Below we summarize this methodology and describe the data and modeling choices specific to our study of the field and globular cluster populations of Fornax.

\begin{table}
\centering
\def\arraystretch{1.5}
\begin{tabular}{||c c c c c||} 
 \hline
 GC & Age & Redshift & $M_{\star}$ at Birth & [Fe/H]\\ & [Gyr] & & [10$^5$M$\odot$] \\& (1) & (2) & (3) & (4)\\ [0.5ex] 
 \hline\hline
 1 & 12.1 $\pm$ 0.8 & $3.69^{+2.49}_{-1.07}$ & 0.42 $\pm$ 0.10 & -2.5 $\pm$ 0.3 \\ 
 2 & 12.2 $\pm$ 1.0 & $3.88^{+4.54}_{-1.07}$ & 1.54 $\pm$ 0.28 & -2.5 $\pm$ 0.3 \\
 3 & 12.3 $\pm$ 1.4 & $4.10^{+26.5}_{-1.83}$ & 4.98 $\pm$ 0.84 & -2.5 $\pm$ 0.3 \\
 4 & 10.2 $\pm$ 1.2 & $1.82^{+0.90}_{-0.51}$ & 0.76 $\pm$ 0.15 & -1.2 $\pm$ 0.1 \\
 5 & 11.5 $\pm$ 1.5 & $2.82^{+3.95}_{-1.11}$ & 1.86 $\pm$ 0.24 & -1.8 $\pm$ 0.2 \\ 
 \hline
\end{tabular}

\caption{Summary of GC properties from \citet{deBoer2016FourDSph}. For each GC, column (1) shows maximum likelhood ages and their respective standard deviation, column (2) shows these ages in redshift space, column (3) corresponds to inferred GC birth masses assuming a kroupa imf, and column (4) shows the present day metallicity.}
\label{table:1}
\end{table}

\subsection{Star Formation History of Fornax's Field Population}

\begin{figure}
\includegraphics[width = \columnwidth]{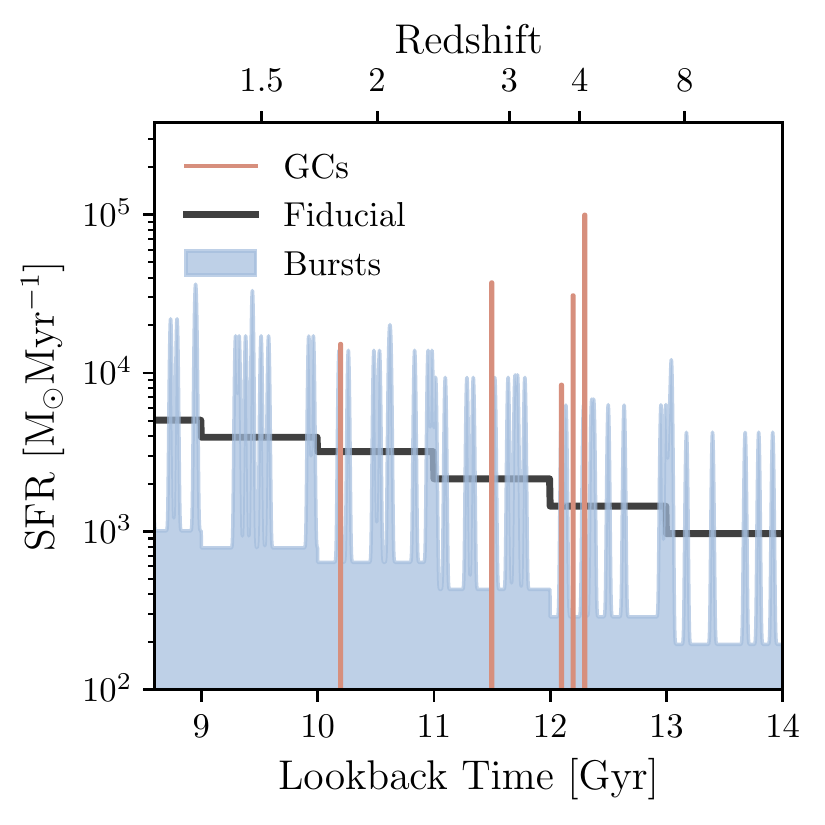}
\caption{\small The combined SFH of Fornax across time. The fiducial tabulated SFH from \citet{deBoer2016FourDSph} is shown in in black, our physically motivated stochastic burst field SFH is overplotted in blue, while GC formation according to the maximum likelihood cluster ages is shown in orange. The GCs essentially act as intense bursts of star formation, amplifying the burstiness of the field.}
\label{fig:SFH}
\end{figure}

For our analysis of Fornax's field population, we use the SFH measured by \citet{deBoer2012TheGalaxy}, which is shown as the solid black line in \autoref{fig:SFH}. This SFH was derived from a deep color-magnitude diagram (CMD) that extends below the oldest main sequence turn-off (MSTO) over $0.8^{\circ}$  (r = $1.9$~kpc) of Fornax's optical body. As described in \citet{deboer2012s}, the SFH was measured with the \texttt{Talos} algorithm, a Kroupa IMF \citep{kroupaimf}, the Dartmouth stellar evolution models \citep{dotter2008} with and age and metallicity range of 0.25 to 15 Gyr and $-2.5$ to $-0.3$ dex, respectively.  An extensive set of artificial stars was used to account for observational uncertainties and crowding.

\subsection{The Ages, Masses, and Metallicities of Fornax's Globular Clusters}

\begin{figure}
\centering
\includegraphics[width = 3.5in]{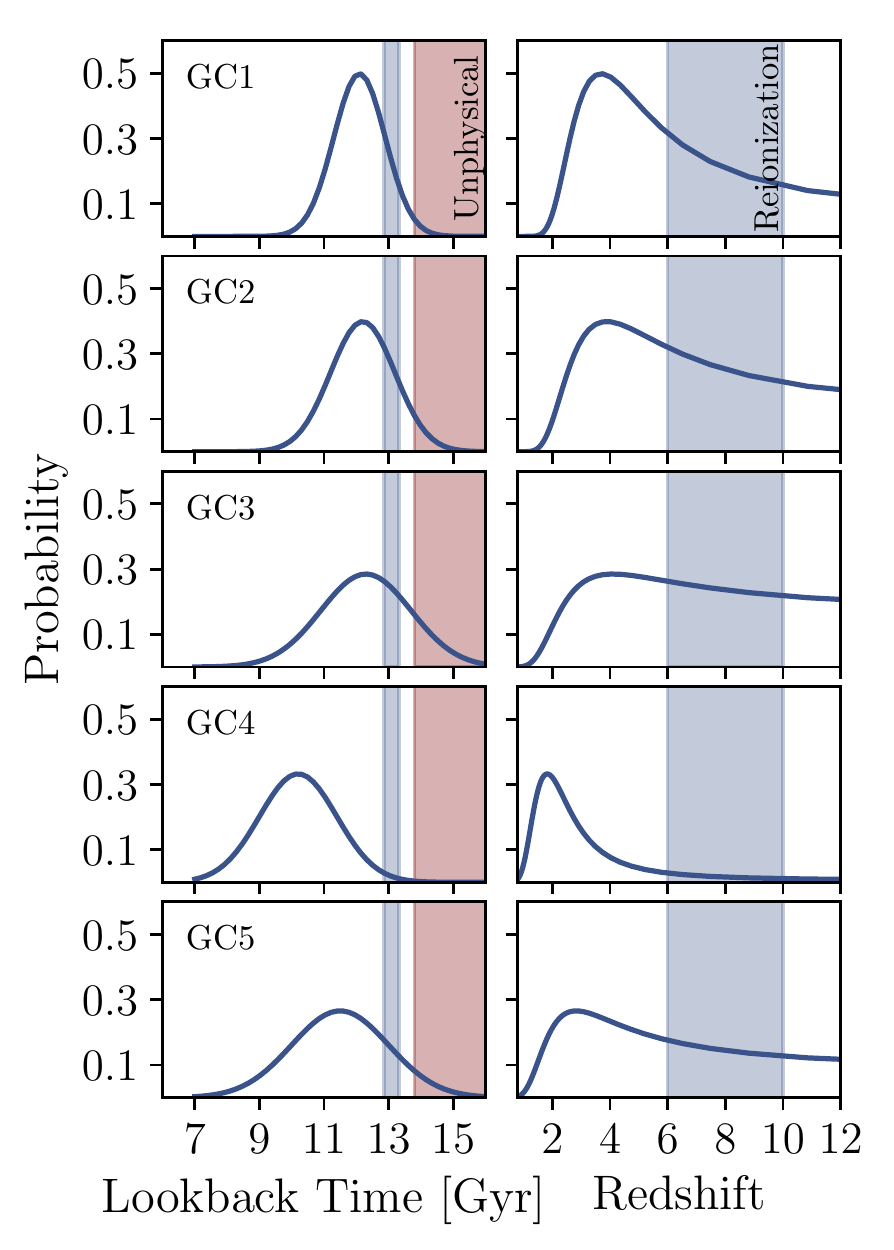}
\caption{\small Age distribution of each Fornax Globular Clusters as a function of lookback time (left) and of redshift (right). The red shaded region denotes lookback times older than the age of the universe according to the latest Planck release. We use this constraint as a prior when deriving distributions in \Muv \ and $\mathrm{\mu_{UV}}$. The blue shaded region corresponds to the epoch of reionization at $6<z<10$, which notably corresponds to only $0.7$~Gyr in lookback time as illustrated on the left panels.}
\label{fig:GC_age}
\end{figure}

We use the ages, masses, and metallicities of Fornax's five GCs from \citet{deBoer2016FourDSph}, which are listed in \autoref{table:1}.  The marginalized age distributions for all five GCs are plotted in \autoref{fig:GC_age}.  

The GC properties were measured using the same analysis techniques as for the field population, ensuring self-consistency. Whereas the field population SFH was measured from ground-based observations, properties of the GCs were derived from CMDs constructed from deep HST/WFPC2 archival imaging. HST imaging was necessary to overcome the high degree of crowding in the GCs and reach the MSTO, ensuring age and metallicity determinations comparable in quality to the field population.

In analyzing Fornax's GCs, \citet{deBoer2016FourDSph} adopt a Kroupa IMF, as opposed to their present day mass function.  As a result, the reported masses reprinted in column (3) of \ref{table:1} are the \emph{birth} masses of the GCs, assuming that they formed with that IMF.  We discuss the role of birth masses in our analysis further in \S \ref{sec:birthmass}.

\subsection{Reconstructing the ultra-violet fluxes of Fornax and its Globular Clusters}\label{SFH}

Following the methodology described in \citet{Weisz2014HUBBLE} and \citet{mbk2015tm}, we reconstruct the rest-frame UV and V-band fluxes of Fornax's field population as a function of redshift using the field SFH (i.e., SFR and metallicity evolution) from the stellar fossil record and the Flexible Stellar Population Synthesis (FSPS) code \citep{Conroy2009THEGALAXIES,Conroy2010THEEVALUATION}. We adopt a Kroupa IMF over a range of .1M$\odot$ to 100M$\odot$, and the Padova stellar evolution models.  Though this is a different stellar library than used for the SFH derivation, the Dartmouth models do not include stars younger than 250 Myr, and thus are not adequate for reconstructing the UV flux from massive, young stars.  

SFHs derived from the stellar fossil record can typically resolve absolute ages to $\sim 10$\% \citep[e.g.,][1 Gyr resolution, 10 Gyr ago]{gallart2005}. However, both observations and simulations indicate that dwarf galaxies have fluctuating SFRs on timescales of $<100$Myr \citep[e.g.,][]{Stinson2007BreathingFormation, ricotti2008, governato2012, power2014, dominguez2015, Onorbe2015ForgedGalaxies}, which affects the UV output from their massive stellar populations.  To account for this effect, we insert a stochastic population of short period bursts into the SFH, following the approach described in \citet{mbk2015tm}. Specifically, we employ a model in which 80\% of star formation occurs in 20 Myr bursts that are 20 times stronger than star formation during the intra-burst period. \autoref{fig:SFH} illustrates our adopted burst scheme (shown in blue) relative to the fiducial SFH of Fornax. Permutations of the burst parameters are explored in \citet{mbk2015tm}, and have minimal impact on the conclusions of this paper. 

Finally, we account for the difference in the areal coverage of the CMD and the entire galaxy. To do this, we assume that the SFH is representative of the entire galaxy, and normalize the modeled present day V-band to the observed value of M$_V = -13.4$ \citep{McConnachie2012TheGroup}. Note that while this absolute magnitude includes the light from the GCs, their combined contribution at the present day is negligibly small ( $\sim1 $\%) compared to the luminosity of Fornax's field population. Finally, we omit the uncertainties on the fiducial SFH as their contribution to the UV flux profiles is negligible compared to the variation introduced by short timescale bursts \citep{Weisz2014HUBBLE}.  We discuss the role of bursts in \S \ref{ml}.

We reconstruct the UV and V-band fluxes of Fornax's GCs self-consistently with the field population (i.e., same IMF, stellar models, FSPS). We first compute a single flux evolution profile across redshift for each GC, using only the most likely combination of age, metallicity, and mass as listed in Table \ref{table:1}. For this reconstruction, we assume a constant SFH over a 5 Myr period, an approximate timescale for star cluster formation \citep[e.g.,][]{lada2003,mckee2007, lada2010, krumholz2015_sf}. This most likely formation scenario is illustrated by the orange lines in \autoref{fig:SFH}. Effectively, under these assumptions, the GCs appear as additional bursts of star formation on top of the field population, where we've assumed no correlation between star formation in the field and GC formation.

\begin{figure*}
\centering
\includegraphics[width = 6.8in]{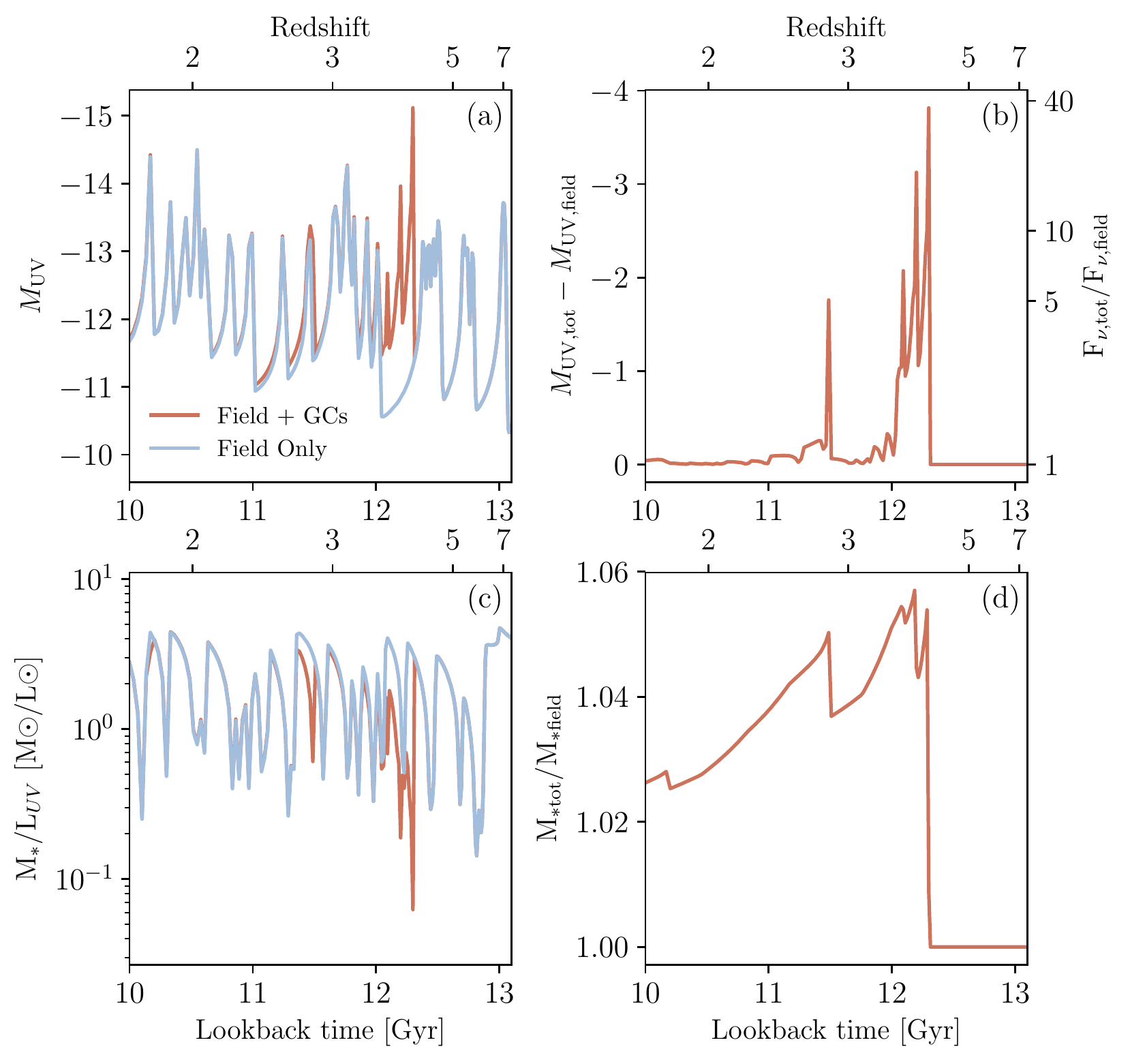}
\caption{\small The reconstructed UV luminosity and stellar mass of Fornax across cosmic time for a standard field SFH. Panel (a): \Muv \ as a function of redshift and lookback time, where the blue line corresponds to the UV magnitude from solely the field and the orange line corresponds to the contribution of GCs assuming the most likely formation time for each of the 5 clusters. The combined field and GC population is at least one magnitude brighter than a typical burst of star formation in the field. This increase allows for substantially more robust measurements in the HFF. Panel (b): The difference in absolute magnitude between the field and the combined field and GC populations. This corresponds to a difference of $\sim4$~mags for this example field SFH scenario. Panel (c): The mass to light (ML) ratio in the UV normalized to the mean field ML ratio, as a function of redshift and lookback time. Though the field ML ratio remains within a magnitude of the mean field ML ratio, proto-GCs can differ by over an order of magnitude. Panel (d): The relative mass of GCs to the field. GCs account for $<5\%$ of the system's mass but can contribute fifty times the UV luminosity.}
\label{fig:mass_flux}
\end{figure*}

As shown in \autoref{fig:GC_age}, the age uncertainties on the GCs are considerable ($\sim 1$ Gyr), particularly when plotted as a function of redshift rather than lookback time. Thus, it is important to also consider the effects of these age uncertainties in our reconstruction of UV fluxes.

We do this using a Monte Carlo process. Assuming that the stellar fossil record provides a Gaussian probability distribution function (PDF) in lookback time with mean and standard deviation as listed in \autoref{table:1}. For each age PDF, we randomly draw a GC birth age, and compute its UV and V-band flux.  We assume the maximum likelihood values for the mass and metallicity, as these are narrowly peaked and do not change considerably with age within the age pdf of a given cluster. We repeat this process $10^4$ times to build up the distribution of \Muv($z$).  We also adopt a prior on age, such that ages allowed by the stellar fossil record, but that exceed the cosmologically derived age of the Universe \citep{Ade2016iPlanck/iResults}, are assigned a probability of zero. We discuss the results of this exercise in \S \ref{age_uncertainties}.

We generate an analogous \Muv($z$) probability distribution function (PDF) for the field by running $10^4$ realizations of the bursty SFH described above for the field population to account for the stochasticity of the short duration bursts. We compute the composite field and GC UV flux profiles by summing the two resulting flux distributions.

Finally, we also derive a PDF for surface brightness by using our \Muv($z$) distributions and adopting sizes for the GC and field populations. For the GCs, we assume an average $r_e$ of 10pc in concordance with sizes of dense bound clusters from simulations \citep{kim2017}. For the field population, we adopted an $r_e = 0.5$ kpc corresponding to values from simulated Fornax-like progenitors  \cite{Ma2017sizes}. We discuss the probabilistic interpretation of our results in \S \ref{prob}.

\section{The UV Luminosity of Fornax across cosmic time}
In this section, we present the reconstructed \Muv\ properties of Fornax.  We first consider the case of a bursty field SFH coupled with the most likely GC ages.  We then factor in uncertainties in the GC ages measured from the stellar fossil record. This division first allows us to illustrate the substantial impact of GCs on Fornax's luminosity modulo complications from the stellar fossil record.  We then fold in the uncertainties to demonstrate how current limitations from the stellar fossil record affect our results.

\subsection{Most Likely Globular Cluster Ages}\label{ml}
\autoref{fig:mass_flux} shows the UV flux evolution of Fornax across cosmic time assuming the bursty SFH described in \S \ref{SFH} and the maximum likelihood age of each GC. As illustrated in blue in panel (a), the short timescale bursts and lulls can change the UV flux of the field population by $\sim 2$~mag, which is identical to the findings in \citet{Weisz2014HUBBLE} and \citet{mbk2015tm}. Unlike these studies, we now also consider the impact of the formation of GCs, shown in orange.  The short (5 Myr), intense periods of star formation that result in formation of proto-GCs, have the same effect on the UV flux as extremely strong bursts of star formation. More concretely, the formation of (proto-)GCs can increase the UV luminosity of Fornax by factors ranging between $\sim10-100$ over periods of a few 10s of Myr depending on star formation in the field. In panel (b) of \autoref{fig:mass_flux}, we show the UV flux ratio of proto-GC to field for an example at the center of this range. See \S \ref{appendix} for limiting examples. This illustration has several implications for the interpretation of objects (nominally assumed to be galaxies) directly detected at high redshift.

Panel (c) of \autoref{fig:mass_flux} illustrates how the bursty field SFH and the formation of GCs affects the ratio of stellar mass to UV luminosity (M/L). The bursty SFH (blue) typically causes fluctuations in the M/L ratio that vary by less than an order of magnitude. In contrast the formation of GCs causes a drop in the M/L ratio that can be larger than an order of magnitude. We quantify this effect further in the next paragraph, and discuss the complications that GC formation may introduce into inferring stellar masses, and in turn halo masses, of high-redshift galaxies in \S \ref{abm}.

Fornax's GC population accounts for $\lesssim5$\% of the total stellar mass of Fornax at an given time as shown in panel (d). Comparing this to panel (b) for the same time interval, GC formation produced up to 50 times more UV luminosity than the field population. In relative terms: despite comprising only 5\% of Fornax's stellar mass at $z \sim 4$, the GCs account for 98\% of total UV flux emitted. 

Interestingly, the temporal clustering of GC formation in Fornax, means that the GCs dominated the UV output of Fornax between $\sim12-12.3$ Gyr ($z = 3.51-4.10$) with the peak of $\Muv = -15.3$~mag. The troughs in  normalized UV flux seen in panel (a) of \autoref{fig:mass_flux} during this period correspond to dimming of the GCs (which happens on the order of 50 Myr) and to a lesser extent star formation in the field. Though the mass fraction of GCs does not vary substantially with each realization of stochastic star formation in the field, the fraction of luminosity contributed by GCs is dependant on the amount of star formation in the field. The values shown here correspond to an average field SFH.

\subsection{Probabilistic Approach}\label{prob}

In the limit of perfect knowledge of GC ages, the above analysis would fully capture the luminosity evolution of Fornax across time. However, uncertainties\footnote{Here we consider uncertainties to be the precision in GC ages.  The issue of absolute ages, i.e., the accuracy in mapping age to redshift, is an equally important, though a much more challenging problem. We discuss these challenges further in \S \ref{age_uncertainties}.}  in ages absolute ages derived from the fossil record are typically of order 10\% of the lookback time, which can be substantial \citep[e.g.,][1 Gyr at 10 Gyr $z = 2^{+1.388}_{-0.719}$]{gallart2005}.

\begin{figure*}
\centering

\includegraphics[width = 6.8in]{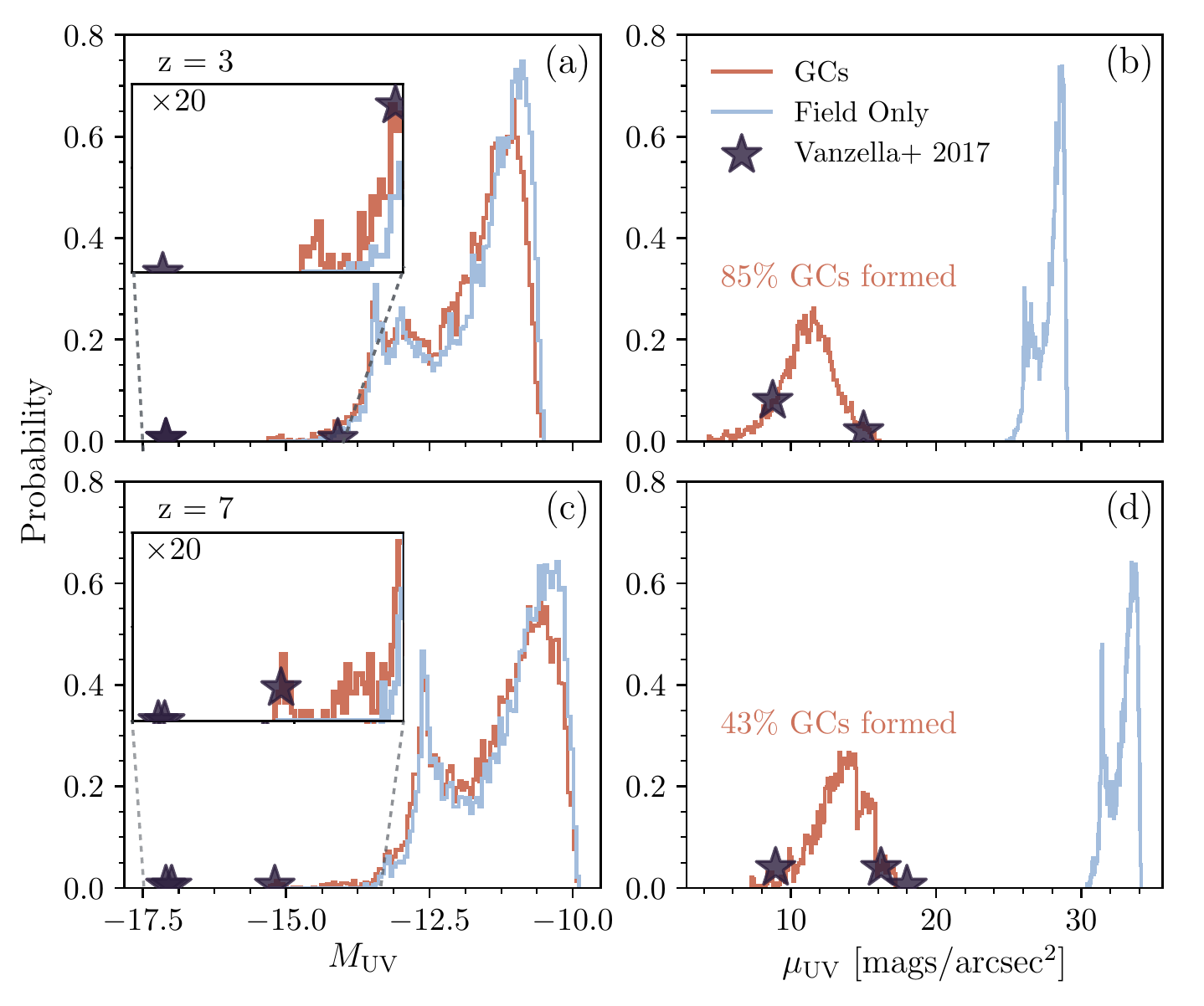}
\caption{\small The probability distribution functions of Fornax and its GCs at select redshifts. Panels (a) and (c): Distribution of $M_{\mathrm{UV}}$ for z = 3 (panel (a)) and z = 7 (panel (c)), where the orange corresponds to the combined GCs+field while the blue corresponds to the field only distribution. In the subpanels of panels (a) and (c) We show a zoom-in (y-axis $\times 20$) on the region of the field + GC PDF corresponding to $2\%$ cumulative probability. At all redshifts the PDF of the combined population is shifted towards more negative magnitudes. This effect is maximized at z$\sim$ 3 when globular clusters are most likely to form. Panels (b) and (d): PDF of $\mu_{UV}$ where here the orange line corresponds to the GC only PDF and the blue still demarcates the field. The two distributions are completely distinct, with the GC PDF falling within the robust detection range of the HFF while the field distribution does not. For both the right and left panels, we overplot the spectroscopically confirmed objects from \citet{Vanzella2017PavingZ3} as purple stars. The objects that fall on the GC distribution in both the \Muv \ and $\mathrm{\mu_{UV}}$ PDFs are most likely to indeed be proto-GCs.}
\label{fig:pdf}
\end{figure*}

To estimate the effects of these age uncertainties on our \Muv\ determinations, we use the Monte Carlo approach described in \S \ref{SFH} to create a probability distribution for the field and GC UV luminosities.  The resulting \Muv\ PDFs are shown in \autoref{fig:pdf} for redshifts $z=3$ (panel (a)) and $z=7$ (panel (c)), which were selected to illustrate the general picture of GC formation near the peak of star formation in the universe ($z\sim2-3$) and during the epoch of reionization ($z\sim7$).

The bimodal distribution seen in both redshift slices is due to the short and bursty star formation in the field.  The more probable faint peak of the distribution (at $\Muv \sim -11.1$ at $z = 3$ and $\Muv \sim -10.5$ at $z = 7$) corresponds to periods between bursts and the less probable peak (at $\Muv \sim -13.7$ at $z = 3$ and $\Muv \sim -12.6$ at $z = 7$) corresponds to stochastic bursts falling within tens of Myr of the considered redshift slice.  

At both redshifts, including GCs only shifts the PDF of the field incrementally towards brighter \Muv. Their main contribution is to add a tail to the bright end of the distribution. At z = 3, this tail corresponds to a maximum $\Muv = -15.5$, which is a magnitude brighter than the field maximum. At $z = 7$ the max $\Muv = -15.3$ and is $3.5$~mag brighter than the field. This smaller offset is due to the lower probability of formation for GCs by $z = 7$ than by $z = 3$. This offset corresponds to a roughly $2\%$ probability that at a given instance, proto-GCs dominate their host's UV luminosity. This is a non-negligible percentage given the ubiquity of Fornax like halos in the Universe. 

As shown in \autoref{fig:pdf}, GCs are more likely to be detected at high redshift than their dwarf galaxies hosts. This is especially relevant for current photometric surveys at these faint magnitudes, as the selection efficiency in the HFF is largely predicated on source size near the surface brightness detection limit \citep{Grazian2011AData}. We therefore examine the surface brightness ($\mathrm{\mu_{UV}}$) probability distribution of the Fornax field compared to its GCs in panels (b) and (d) of \autoref{fig:pdf} for $z = 3$ and $z = 7$, respectively. At both redshift slices, the two distributions are completely distinct and are separated by $\sim 14 \ mags/arcsec^2$. There is an 85\% likelihood that at least one GC has formed by $z = 3$, whereas at $z = 7$ this is lower at 43\%. The slightly higher GC surface brightness possible at $z = 3$ is due to this higher probability of formation and the corresponding increased likelihood of detecting a proto-GC at its most luminous. Finally, in \autoref{fig:pdf} we plot compact objects spectroscopically followed up by \citet{Vanzella2017PavingZ3} along with our PDFs. We comment on their remarkable similarities in \S \ref{GC_HFF}.

\begin{figure*}
\centering
\includegraphics[width = 3.4in]{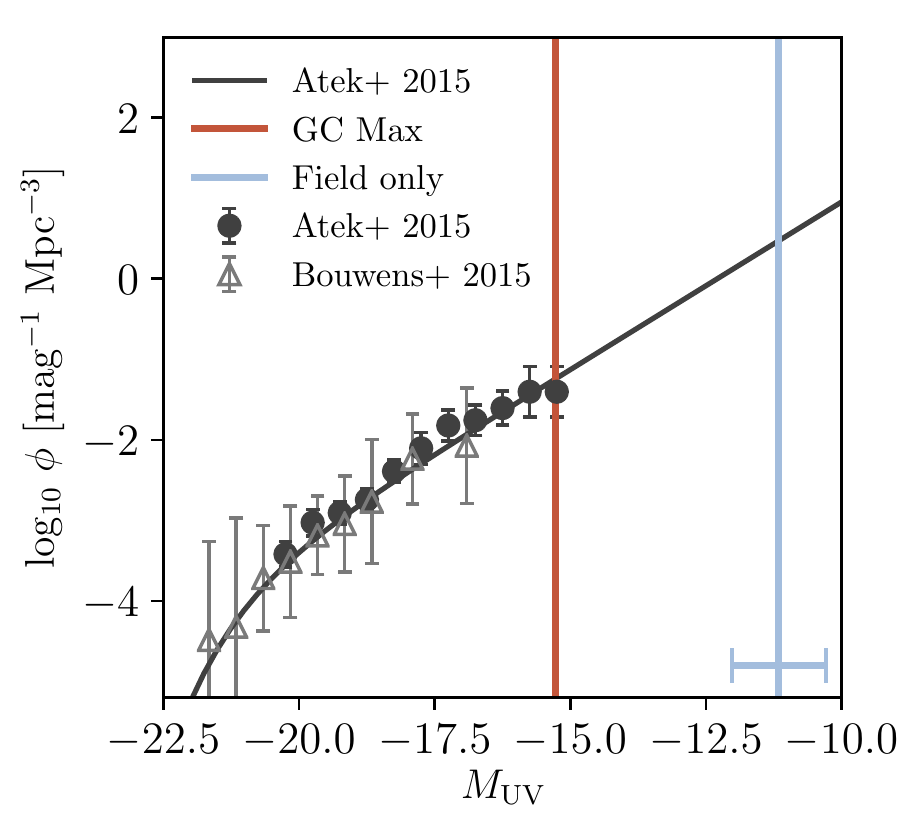}
\includegraphics[width = 3.4in]{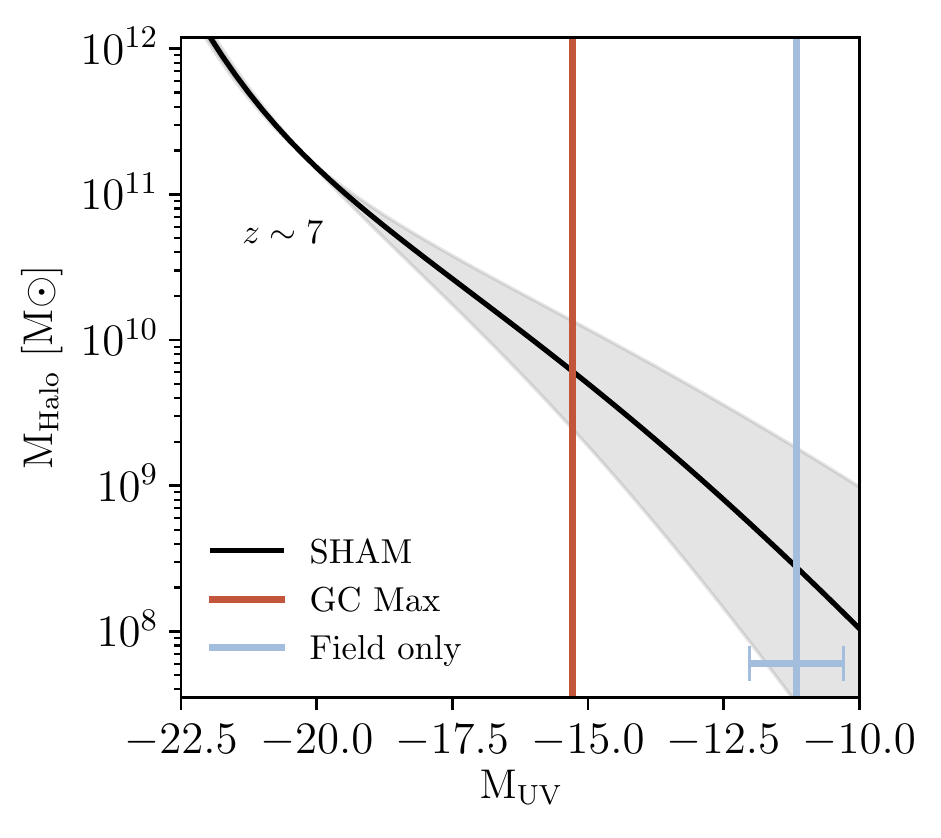}
\caption{\small Illustration of the impact of GC formation on the UVLF and abundance matching at high redshift. On left: comparison of our modeled magnitudes with the luminosity function of $z \sim 7$. We show the best fit UVLF and measured points from \citet{Atek2015AREPARALLELS} in black, as well as the measurements from \citet{Bouwens2015UVFIELDS} in grey. The maximum GC+Field magnitude, shown in orange falls within the detection range of \citet{Atek2015AREPARALLELS}, while the Fornax field on its own, shown in blue, does not. On right: M$_{\mathrm{\mathrm{Halo}}}$ as a function of \Muv. In black, we show a SHAM relation derived from \citet{Sheth2001EllipsoidalHaloes} and \citet{Finkelstein2016Observational6}, where the grey region corresponds to variations of $\pm 0.3$ in the faint end slope of the UVLF. In blue and orange we show where the Fornax field and its proto-GCs would respectively fall on this relation. Abundance matching using the \Muv \ from the field is consistent with the halo mass inferred for Fornax, whereas accounting for GCs can overestimate this by an order of magnitude.} 
\label{fig:lumfunc}
\end{figure*}
\section{Discussion}

\subsection{Proto-Globular Clusters in the Hubble Frontier Fields}\label{GC_HFF}
Gravitational lensing is a powerful means of detecting faint objects in the high-redshift Universe \citep[e.g.,][]{Atek2015AREPARALLELS, Alavi2016THE3,livermore2017, Bouwens2017ExtremelyEmissivity}.  However,  the nature of shear and flux amplification from lensing, means that the HFF yields preferential detection of compact, rather than extended sources, for a given intrinsic magnitude and magnification parameter \citep[e.g.,][]{Grazian2011AData, wong2012, Oesch2015, Atek2015AREPARALLELS, Alavi2016THE3, Bouwens2017ExtremelyEmissivity}. 

\autoref{fig:pdf} illustrates this effect for Fornax. Both the field and GCs are near the detection UV flux limit of the HFF \citep{Grazian2011AData}, but the PDF for GCs at a both $z = 7$ and $z = 3$ is centered at a substantially higher surface brightness than the field.  This is simply due to their relative sizes. Thus, at faint magnitudes in the HFF, it is substantially more likely to detect a GC in Fornax, than Fornax itself. Based on this analysis, we caution that some of the faintest objects detected at high redshift \citep[e.g., $\Muv = -12.5$ at $z = 7$;][]{livermore2017} to date may in fact be GCs hosted by faint galaxies, and not faint galaxies themselves.

This conjecture is consistent with recent demonstrations that faint objects in the HFF are generally quite compact \citep[e.g.,][]{Kawamata2015THEDATA,Laporte2016}. For example, \citet{Bouwens2017ExtremelyEmissivity} show a large population of faint ($\Muv\leq-16$) fall between $15 \ \mathrm{pc}<r_e<55 \ \mathrm{pc}$ at $z \sim 6$. While we find both the Fornax field and GCs to potentially have UV magnitudes on par with these objects, the surface brightness we infer for its GCs at $z = 3$ and $z=7$ are more consistent with these results than the field population alone. 

\autoref{fig:pdf} also compares the spectroscopically confirmed compact objects detailed in \citet{Vanzella2017PavingZ3} to our findings for Fornax. Interestingly, all of the high-redshift objects are consistent with the GC surface brightness distribution, but not the field population distribution. Notably, both the proto-GC candidates fall on the \Muv \ distribution for Fornax GCs as well. 

Moreover, the high-redshift objects and Fornax GC have similar physical properties.  For example, Fornax GC3 has a mass of $\sim 5\times 10^5$ M$_{\odot}$, which corresponds to an SFR of $\sim 0.1M_{\odot}yr^{-1}$ and a maximum \Muv \ of $-15.2$. The most comparable high-redshift object reported by \citet{Vanzella2017PavingZ3} is GC1 at $z = 6.145$, which has a stellar mass of $0.8-130 \times 10^6$ M$_{\odot}$, an SFR between $0.04-6.6M_{\odot}yr^{-1}$, \Muv\ $= -15.3$ mag. The similarity of these values suggests that GC1 ($r_e \sim 10$~pc), may be a proto-globular cluster, though the uncertainties in the SFR and mass are large.  More generally, this type of comparison reinforces the conclusions of \citet{Vanzella2017PavingZ3}, i.e., that they have observed star clusters, and strengthens the potential for connections between compact objects found in the local Universe and those at higher redshifts.

\subsection{Globular Clusters, the UV Luminosity Function, and Abundance Matching}\label{abm}

The left panel of \autoref{fig:lumfunc} illustrates how the GC formation can affect Fornax's position on the $z=7$ UVLF. The broader implication of this GC-driven brightening is that, at least in part, GCs could be counted as galaxies in current UVLF determinations. That is, if GCs from fainter, more numerous galaxies are being counted as more luminous galaxies on their own, it not only biases the UVLF, it also washes out potential structure at the faint end. This could inhibit surveys from detecting the turnover in the UVLF, which is expected from a number of detailed simulations of high-redshift galaxy populations \citep[e.g.,][]{jaacks2013, oshea2015, gnedin2016, finlator2017}, as well as consistency with the stellar fossil record and number counts of Local Group galaxies \citep[e.g.,][]{mbk2014, mbk2016, Weisz2017LocalEra}.

Intriguingly, \citet{bouwens_uncertainties} indicates that there may be a tension between the faint-end UVLF slope from the HFF versus HUDF. Given differences in the selection effects of lensed vs. blank fields (as discussed in \S \ref{GC_HFF}), this may be a sign that proto-GCs are making non-negligible contributions to the UVLF. Broader implications of how GC-galaxy confusion at high-redshifts affects reionization and the UVLF have been discussed at several places in the literature \citep[see][]{mbk2017theconnection, mbk2017theformation, Bouwens2017complexes}.

Proto-globular cluster formation may also affect abundance matching relationships. To investigate the implications of this, we calculate the stellar-halo abundance mass relation (SHAM) given the \citet{Sheth2001EllipsoidalHaloes} halo mass function and the $z = 7$ UVLF from \citet{Finkelstein2016Observational6}, varying the fiducial faint end slope over $\Delta \alpha = \pm 0.3$ to encapsulate the proposed values in the literature \citep[e.g,][]{Stark2016GalaxiesBang}. 

The right panel of \autoref{fig:lumfunc} shows that for the case of Fornax at $z\sim7$, if its UV luminosity is dominated by a proto-GC, then the assigned halo mass would be over-estimated by a factor of $\sim20$ given our fiducial faint end slope. That is, the UV luminosity of the field population alone would correspond to a halo mass of $3 \times 10^8$~M$_{\odot}$, whereas including the luminosity boost from GC formation would imply a halo mass of $6 \times 10^9$~M$_{\odot}$. This increases to a factor of $\sim70$ if we consider a steeper faint end slope as some studies suggest \citep{livermore2017, ishigaki2017}. Though the exact difference will depend on the adopted SHAM, the order-of-magnitude discrepancy for the fiducial case is approximately correct, as all SHAMs are similarly steep. Beyond revealing a shortcoming in abundance matching at high redshifts, this mis-assignment of halo masses has a range of implications ranging from incorrectly interpreting the astrophysics of faint UV sources to differentiating between dark matter models, which can predict different shapes to the faint end of the UVLF \citep[e.g.,][]{ Schultz2014TheMatter, Bozek2015GalaxyMatter, Menci2017FundamentalGalaxies}.

\subsection{Globular Clusters in Other Local Group Dwarfs}
As the Fornax field approaches the detection limit of the HFF, we can leverage our results to predict the high-redshift observability of fainter and brighter GC hosting dwarfs. In fainter hosts, GCs can act as tracers of galaxies that would be beyond the detection limits of present or future surveys. There are four local group dwarfs fainter than Fornax that host a star cluster: PegDIG \citep[$M_{\mathrm{v}} = -12.2$;][]{Cole2017DDOGalaxy}, AndI \citep[$M_{\mathrm{v}}$ = -11.7;][]{Cusano2016VariableNucleus}, AndXXV \citep[$M_{\mathrm{v}}$ = -9.7;][]{Cusano2016VariableNucleus}, and Eridanus II \citep[$M_{\mathrm{v}}$ = -7.1;][]{Crnojevic2016DEEPCLUSTER}. With the possible exception of PegDIG, the progenitors of these galaxies will not be observable at redshifts relevant to reionization, even with \textit{JWST} \citep{mbk2015tm}. By the surface brightness projections made in this paper, their GCs may however be detected. This could be used to constrain the number density of extremely faint galaxies and inform models of reionization \citep[e.g.,][]{Robertson2013NewCampaign, Robertson2015CosmicTelescope}.

The progenitors of GC-hosting LG dwarfs brighter than Fornax, like WLM \citep[$M_{\mathrm{v}} = -14.2$][]{Leaman2012TheWLM}, NGC 6822 \citep[$M_{\mathrm{v}} = -15.2$][]{Hwang2011Extended6822, Huxor2013}, the LMC and SMC \citep[e.g.,][]{Forbes2015TheClusters}, could fall within the robust surface brightness detection limits of the HFF. This means one could simultaneously observe the host and its proto-GCs at high redshift, informing our picture of high-redshift star formation. 

\subsection{Next Steps: Building Connections Between Local Globular Clusters and High-Redshift Compact Objects}
\label{sec:future}

Up to this point, we have used LG dwarf galaxy Fornax and its GC population to illustrate a fundamental connection between the stellar fossil record of local systems with high-redshift observations. Given such studies are in nascent stages, we now highlight a few ways in which local and high-redshift studies of clusters can be strengthened.

\subsubsection{The Impact of Globular Cluster Birth Mass}\label{sec:birthmass}

One challenge in connecting GCs observed in the local Universe with putative progenitor populations at high redshifts, lies with their stellar masses. It is well-established that low-mass stars have been ejected from GCs over their lifetimes due to dynamical interactions within the dense cluster environment \citep[e.g.,][]{Ostriker1972OnClusters, Chernoff1986TidalClusters, gnedin1999, fall2001}.

The effect of this `evaporation' is that GCs today are likely to be less massive that when they formed.  In turn, the more massive a GC was when it formed, the brighter it would have been (assuming a Galactic-like stellar IMF).  Furthermore, most theoretical explanations for the presence of multiple populations in MW GCs require that they formed with significantly larger stellar masses \citep[factors of 10-100; e.g.,][]{Piotto2012iHUBBLECLUSTERS, Renzini2015TheScenarios}. However, see \citet{bastian2017} for claims that these scenarios are not physically viable.

In the case of Fornax, there are indications that its GCs have lost no more than a factor of few in stellar mass over their lifetimes.  Comparisons between the metal-poor star population in the field of Fornax and the GC metallicities suggest that cluster stellar mass loss is no more than a factor of 4-5 \citep{Larsen2012ConstraintsDSph, deBoer2016FourDSph}.  

In this paper, we have used estimates of GC {\emph{birth}} mass for our analysis.  In modeling the CMDs of Fornax's GC population, \citet{deBoer2016FourDSph} assumed a Kroupa IMF, which would correct for the mass loss affect, under the assumption that the GCs formed with that IMF. However, if the assumption of a Kroupa IMF is not correct \citep{Zaritsky2012EVIDENCEFUNCTIONS} and/or the birth masses of Fornax GCs were larger than we have assumed, Fornax's GC may have been even more UV-luminous than we find.

\subsubsection{The Role of Absolute Age Uncertainties}\label{age_uncertainties}

In our analysis, we have assumed a specific mapping between lookback time and redshift.  However, it is well-established that stellar and GC absolute ages are uncertain and depend on the detailed stellar physics \citep[see][and references therein]{soderblom2010}. This introduces challenges into translating ages from the fossil record into a cosmological reference frame (i.e., redshift).  

\autoref{fig:GC_age} illustrates this issue. Here, the inferred age distribution as a function of lookback time is Gaussian.  However, in terms of redshift a significant portion of the PDF extends to ages older than allowed by cosmological models. Improving knowledge of absolute ages requires better observational \citep[e.g. absolute distances; ][]{Vandenberg2013THEISSUES, Chaboyer2017TestingStars,OMalley2017} and theoretical underpinning \citep[e.g. stellar physics; ][]{Bonaca2012CalibratingView, Tanner2014TheModels,Creevey2014BenchmarkData}.   

A related issue is the precision to which GC ages can be measured. In the case of Fornax, \citet{deBoer2016FourDSph} report GC ages to a precision of $\pm1$~Gyr.  As \autoref{fig:GC_age} shows, this translates into considerable uncertainty on the redshift distribution.  Even if absolute ages were not an issue, determining whether a given GC formed before/during/after the epoch of reionization is challenging simply owing to precision.  As illustrated in Figure \autoref{fig:GC_age}, the epoch of reionization is $\sim0.7$ Gyr in duration, and sets a requirement on the precision to which GC ages from the stellar fossil record must be known to determine their relationship to reionization.  

There are several avenues that should improve the precision, and possibly accuracy, to which GC ages can be measured.  First, \textit{Gaia} \citep{Lindegren2016GaiaParallaxes}, will provide distance measurements to galactic GCs with a precision of $\approx 1\%$, a factor of $>10$ improvement over most distance estimate to date.  Such precise parallaxes should limit GC age precision to no less than 10\% \citep{Pancino2017GlobularGaia}.  Second, the accessibility of the `MS kink', a feature in the low-mass portion of the CMD caused by changes in atmospheric opacity, may improve age precision. The MS kink is metallicity sensitive, and could mitigate the age-metallicity degeneracy that affects measuring GC properties from the MSTO \citep{Correnti2016CONSTRAININGDIAGRAM}. Determining absolute GC ages is a far more challenging problem as it requires an improved understanding of detailed stellar physics \citep[e.g.,][]{Vandenberg2013THEISSUES, Chaboyer2017TestingStars, Tayar2017}.

\subsubsection{Detecting proto-Globular Clusters at High Redshifts}\label{future_obs}

Based on arguments advanced in this paper and elsewhere, it appears that proto-GC are likely already being detected at high redshifts in the HFF. Future spectroscopic follow up of these sources may help confirm their nature as proto-GC through determinations of stellar and dynamical mass, specific star formation rates, and secure redshifts, which can substantially improve size determinations \citep[e.g.,][ and references therein]{treu2015}.  
 
Unfortunately, ancient metal-poor GCs like four of the ones found in Fornax, should not have a detectable Ciii] doublet, which is the brightest spectral signature in the restframe UV after Ly$\alpha$ \citep{Stark2014UltravioletZgt7}. Line emission from the Ciii] doublet peaks at a metallicity of $\mathrm{logZ} = -0.7$ \citep{Erb2010PHYSICAL} and becomes more difficult to detect for metallicities lower than $\mathrm{logZ} = -1.5$, even for sources with high ionization parameters \citep{Jaskot2016PHOTOIONIZATIONGALAXIES}.
 
One possible way to uniquely detect proto-GCs at high redshift may arise from the multiple chemically distinct populations found in Galactic clusters \citep[e.g.,][]{gratton2012, Piotto2012iHUBBLECLUSTERS}.  The so-called `second generation' of stars exhibits distinct abundance patterns such as Helium enhancement and anti-correlations between oxygen and sodium \citep[e.g.,][]{Villanova2012THE, Vandenberg2013THEISSUES, bastian2017}. Presumably, such unusual chemical signatures could be seen in emission as the proto-GC were forming.

\section{Summary}
In this work, we reconstructed the UV luminosity of the Fornax field and its GCs using their respective stellar fossil records combined with stellar population synthesis modeling. We have shown that forming GCs can be substantially brighter than their dwarf galaxy hosts at high redshift. Specifically, we find: 
\begin{enumerate}[(i)]
    \item Proto-GCs can emit 50 times the UV luminosity of their dwarf galaxy hosts despite comprising $\leq 5\%$ of the total mass.
    \item Forming GCs can be brighter than $\Muv = -15$, which lies in the robust detection limit of the HFF at $z \gtrsim 6$. 
    \item GCs are described by a substantially higher surface brightness PDF than their dwarf galaxy hosts.
\end{enumerate}
Given that both Fornax and its GCs are near the detection UV flux limit of the HFF, the higher surface brightness of GCs means they are more likely to be detected than a dwarf galaxy that hosts them. We further assess the implications of preferentially detecting clusters and find: 
\begin{enumerate}[(i)]
    \item Miscategorizing GCs as galaxies at high-redshift could bias the UVLF and wash out signatures of a turnover at the faint end.
    \item Inferred halo masses from abundance matching relations could be more than an order of magnitude too massive if a proto-GC is mistaken for its host dwarf galaxy.
\end{enumerate}

With improvements in age precision from \textit{Gaia} and enhanced detection limits and spectroscopy from \textit{JWST}, synthesizing low and high redshift observations is a promising avenue for understanding the formation of GCs and their role in the early universe. 

\section*{Acknowledgements}

The authors would like to thank Thomas de Boer for making the SFH of Fornax and its GCs available to us. This work was performed under the auspices of the U.S. Department of Energy by Lawrence Livermore National Laboratory under Contract DE-AC52-07NA27344. Funding for TOZ was provided by LLNL Livermore Graduate Scholar Program. DRW is supported by a fellowship from the Alfred P. Sloan Foundation. MBK acknowledges support from NSF grant AST-1517226 and from NASA grants NNX17AG29G and HST-AR-13888, HST-AR-13896, HST-AR-14282, HST-AR-14554, HST-AR-15006, HST-GO-12914, and HST-GO-14191 from the Space Telescope Science Institute, which is operated by AURA, Inc., under NASA contract NAS5-26555. This paper made use of \texttt{iPython} \citep{ipython} and the \texttt{PYTHON} packages \texttt{NUMPY} \citep{numpy}, \texttt{MATPLOTLIB} \citep{matplotlib}. This research has made extensive use of NASA's Astrophysics Data System and the arXiv eprint service at arxiv.org.

\bibliographystyle{mnras}
\DeclareRobustCommand{\De}[3]{#3}
\bibliography{mendeley2,references.bib}
\clearpage
\appendix
\section{GC Formation with Varying Field SFH}\label{appendix}
\begin{figure*}
\centering
\includegraphics[width = 6.8in]{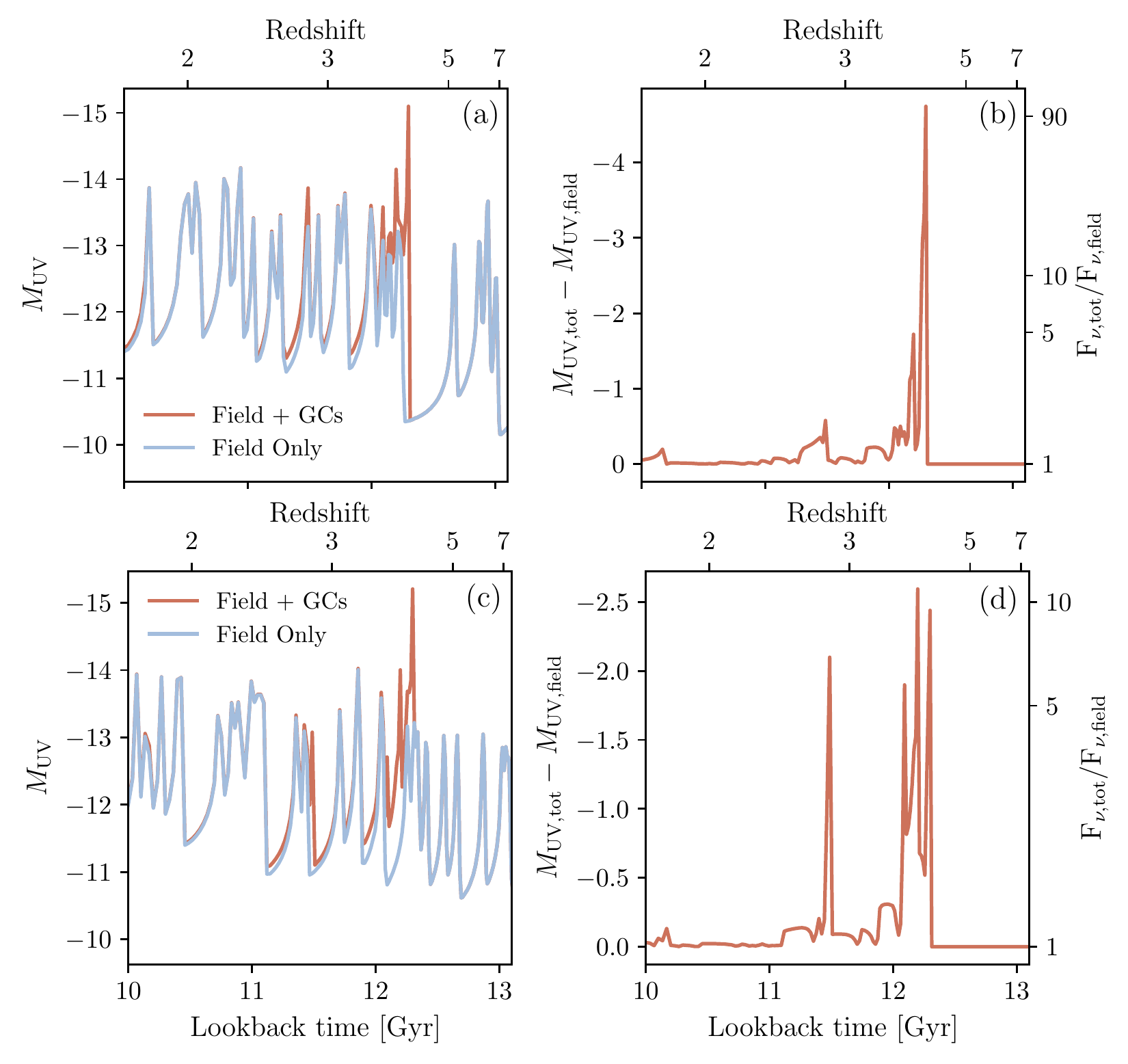}
\caption{\small The reconstructed UV luminosity of Fornax across cosmic time for extreme cases of field star formation with respect to GC formation. Panel (a): \Muv \ as a function of redshift and lookback time, where the blue line corresponds to the UV magnitude from solely the field and the orange line corresponds to the contribution of GCs assuming the most likely formation time for each of the 5 clusters. The formation of GCs corresponds to a lull in star formation in the field. Panel (b): The difference in absolute magnitude between the field and the combined field and GC populations. This corresponds to a difference of $\sim5$~mags for this low field star formation scenario. Panel (c): Shows the same quantities displayed in panel (a) but for a field SFH corresponding to a simultaneous burst in galaxy star formation and GC formation. Panel (d): For this case, the maximum difference in magnitude between proto-GCs and the field is 2.5 mags, this means forming GCs are still ten times more luminous in the UV than the burst in the field.}
\label{app}
\end{figure*}

The UV flux ratio between proto-GCs and the field for the Fornax dwarf spheroidal ranges between $\sim10-100$ depending on the relationship between star formation in the field and GC formation. We explicitly illustrate the effect of this relationship on the UV luminosity breakdown of Fornax in \autoref{app}. We show two limiting cases: panels (a) and (b) correspond to the limit were GC formation is coincident with a lull in field star formation, while panels (c) and (d) show GCs forming at a peak of star formation in the field.

Hydrodynamical simulations show clusters are likely to form when there is an enhancement of cold gas reservoirs in the galaxy \citep{kravtsov2005}; however, this does not necessarily correspond to enhanced star formation in the field especially for metal-poor GCs (i.e.: four of the Fornax GCs). Generally these are thought to form either earlier than the primary epoch of star formation in the galaxy\citep[e.g.,][]{forbes1997}, from dissipational collapse at the center of low-mass halos in assembly based models of GC formation \citep[e.g.,][]{tonini2013}, or in high-redshift merger events that require only enough cold gas to form a GC (i.e: $>10^5$M$_{\odot}$) \citep[e.g.,][]{li2014}. Depending on the GC formation mechanisms at play within Fornax, we would expect varying degrees of correlation between star formation in the field and GC formation. The case in which GC formation is highly correlated with star formation in the field is shown in panels (c) and (d) of \autoref{app}.

\bsp	
\label{lastpage}
\end{document}